\documentclass[twocolumn,aps,pra,showpacs]{revtex4}
\usepackage{epsfig}
\begin{document}
\title{Knowledge excess duality and violation of Bell inequalities}
\author{Radim Filip and Miroslav Gavenda}
\affiliation{Department of Optics, Palack\' y University,\\
17. listopadu 50,  772~07 Olomouc, \\ Czech Republic}
\date{\today}
\begin{abstract}
A constraint on two complementary knowledge excesses by maximal
violation of Bell inequalities for a single copy of any mixed
state of two qubits $S,M$ is analyzed. The complementary knowledge
excesses ${\bf \Delta K}(\Pi_{M}\rightarrow \Pi_{S})$ and ${\bf
\Delta K}(\Pi'_{M}\rightarrow \Pi'_{S})$ quantify an enhancement
of ability to predict results of the complementary projective
measurements $\Pi_{S},\Pi'_{S}$ on the qubit $S$ from the
projective measurements $\Pi_{M},\Pi'_{M}$ performed on the qubit
$M$. For any state $\rho_{SM}$ and for arbitrary
$\Pi_{S},\Pi'_{S}$ and $\Pi_{M},\Pi'_{M}$, the knowledge excesses
satisfy the following inequality ${\bf \Delta
K}^{2}(\Pi_{M}\rightarrow \Pi_{S})+{\bf \Delta K}^{2}
(\Pi'_{M}\rightarrow \Pi'_{S})\leq (B_{\mbox{max}}/2)^2$, where
$B_{\mbox{max}}$ is maximum of violation of Bell inequalities
under single-copy local operations (local filtering and unitary
transformations). Particularly, for the Bell-diagonal states only
an appropriate choice of the measurements $\Pi_{S},\Pi'_{S}$ and
$\Pi_{M},\Pi'_{M}$ are sufficient to saturate the inequality.
\end{abstract}
\maketitle

\section{Introduction}

Immediately after the discovery of quantum mechanics, it was
realized that it contains an interesting feature in quantum
correlations between two particles. It was first discussed in
seminal paper of Einstein, Podolsky and Rosen (EPR) for the
coordinate and momentum of a pair of massive particles \cite{EPR}
and after a time reformulated for spin-entangled systems
\cite{Bohm51}. Assuming a pair of maximally entangled spin-1/2
particles we can perfectly predict the results of the
complementary measurements on one particle from an appropriate
measurements of the other one. An ability of the precise
prediction of complementary variables arises from quantum nature
of the correlations between the particles. From a fundamental
point of view it was proved that for such the particles the
measurements of correlated quantities should yield a different
result in the quantum mechanical case to those expected in local
realism. A condition derived in a form of Bell inequality has to
be satisfied within the local realism \cite{Bell64}. The
predictions of quantum mechanics were satisfactorily
experimentally proved using pairs of photons entangled in the
polarization \cite{Kwiat95,Weihs98}. From a practical point of
view, such the entangled particles distributed at a distance can
be used to securely distribute classical information.

An experiment demonstrating the interesting attribute of the
correlations between quantum systems can be build up assuming a
generally mixed state $\rho_{SM}$ of qubit $S$ and meter qubit
$M$. This experiment is schematically depicted in Fig.~1.
Performing two projective (ideal) measurements $\Pi_{M},\Pi'_{M}$
on the qubit $M$, the prediction of the results of the
complementary measurements $\Pi_{S},\Pi'_{S}$ on the qubit $S$ can
be improved. The complementarity of the measurements means that
$\mbox{Tr}\Pi_{S}\Pi'_{S}=1/2$. For example, having maximally
entangled state
$|\Psi_{-}\rangle_{SM}=(|VH\rangle-|HV\rangle)/\sqrt{2}$, 
the results of arbitrary measurement $\Pi$ on the qubit
$S$ can be precisely predicted performing the same measurement $\Pi$ on the
qubit $M$. However, without this measurement we have vanishing
knowledge since the state of $S$ is maximally random. The total
ability to predict the result of measurement was quantified by a concept of 
{\em knowledge} defined in \cite{Englert}. Generally, there are such
states for which the both complementary knowledges ${\bf
K}(\Pi_{M}\rightarrow \Pi_{S})$ and ${\bf K}(\Pi'_{M}\rightarrow
\Pi'_{S})$ obtained from the measurements $\Pi_{M},\Pi'_{M}$ are
larger than these ${\bf P}(\Pi_{S})$ and ${\bf P}(\Pi'_{S})$
without the measurements. Then corresponding knowledge excesses
${\bf \Delta K}(\Pi_{M}\rightarrow \Pi_{S})={\bf
K}(\Pi_{M}\rightarrow \Pi_{S})-{\bf P}(\Pi_{S})$ and ${\bf \Delta
K}(\Pi'_{M}\rightarrow \Pi'_{S})={\bf K}(\Pi'_{M}\rightarrow
\Pi'_{S})-{\bf P}(\Pi'_{S})$ can be introduced, respectively for
the complementary measurements. A duality between the knowledge
excesses for any mixed state $\rho_{SM}$ can be derived,
analogically as in Ref.~\cite{Englert}. For $\Pi_{M}=\Pi'_{M}$,
the knowledge excesses satisfy the inequality ${\bf \Delta
K}^{2}(\Pi_{M}\rightarrow \Pi_{S})+{\bf \Delta
K}^{2}(\Pi_{M}\rightarrow \Pi'_{S})\leq 1$. Thus performing a
single measurement on $M$, the sum of the squares of knowledge
excesses cannot be larger than unity.

In this paper, we analyze a duality between knowledge excesses
beyond the condition $\Pi_{M}=\Pi'_{M}$ and derive the following
restriction of the complementary knowledge excesses ${\bf \Delta
K}(\Pi_{M}\rightarrow \Pi_{S})$ and ${\bf \Delta
K}(\Pi'_{M}\rightarrow \Pi'_{S})$, namely
\begin{eqnarray}\label{main1}
{\bf \Delta K}^{2}(\Pi_{M}\rightarrow \Pi_{S})+{\bf \Delta K}^{2}(\Pi'_{M}\rightarrow \Pi'_{S})
\leq\left[\frac{B_{\mbox{max}}}{2}\right]^{2},
\end{eqnarray}
where $B_{\mbox{max}}$ is maximal violation of Bell inequalities
\cite{Horodecki95,Zukowski01} after optimal local filtering
operations on a single copy of  the state $\rho_{SM}$
\cite{Verstraete02}. Thus  $B_{\mbox{max}}$ restricts the ability
to enhance both the knowledge excesses by the different measurements on
$M$. To overcome the unit value of sum of squares of the knowledge excesses we
need a state violating the Bell inequality. For any state having
vanishing the a priori knowledge in any basis (for example, Bell-diagonal
state) we achieve the equality only by choosing appropriate
measurements $\Pi_{S},\Pi'_{S}$ and $\Pi_{M},\Pi'_{M}$. Such the
state, exhibiting the maximal accessible $B_{\mbox{max}}$ under
local filtering on a single copy and unitary operations, can be
probabilistically but uniquely obtained from an arbitrary
two-qubit state using local single-copy filtrations
\cite{Verstraete02}. Thus the inequality (\ref{main1}) with
$B_{\mbox{max}}$ can be always saturated assuming an appropriate
local filtration before the suitable measurements
$\Pi_{S},\Pi'_{S}$ and $\Pi_{M},\Pi'_{M}$. We analyze simple and
experimentally feasible examples covering such the cases in which
noise prevents an maximal extraction of knowledge.

\begin{figure}
\centerline{\psfig{width=8.0cm,angle=0,file=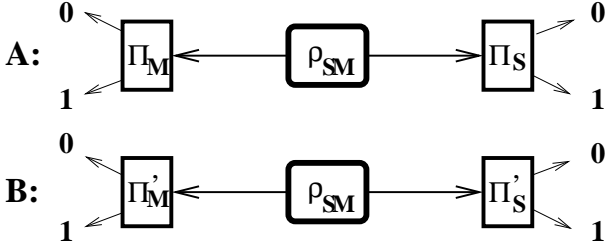}}
\caption{The experiment.}
\end{figure}

\section{Knowledge excess duality}

In this Section we define the complementary knowledge excesses
${\bf \Delta K}(\Pi_{M}\rightarrow \Pi_{S})$ and ${\bf \Delta
K}(\Pi'_{M}\rightarrow \Pi'_{S})$ and discuss a duality between
them which arises from a single measurement on $M$. Assume
two-component projective measurement
$\Pi_{S}=\{|\Psi\rangle_{S}\langle\Psi|,|\Psi^{\bot}\rangle_{S}\langle\Psi^{\bot}|\}$
giving results either $0$ or $1$, respectively. We expand the
state $\rho_{SM}$ as
\begin{eqnarray}\label{state}
\rho_{SM}=w|\Psi\rangle_{S}\langle \Psi|\rho_{M}+
w^{\bot}|\Psi^{\bot}\rangle_{S}\langle\Psi^{\bot}|\rho_{M}^{\bot}+\nonumber\\
\sqrt{ww^{\bot}}\left(|\Psi\rangle_{S}\langle \Psi^{\bot}
|\chi_{M}+\mbox{h.c.}\right),
\end{eqnarray}
where $0\leq w,w^{\bot}\leq 1$, $ w+w^{\bot}=1$ and the meter
operators $\rho_{M},\rho_{M}^{\bot},\chi_{M}$ depend on a choice
of the measurement $\Pi_{S}$. To predict a result of $\Pi_{S}$ on $S$ we can unambiguously discriminate the mixed states
$\rho_{M},\rho_{M}^{\bot}$ by a projective two-component
measurement $\Pi_{M}=\{\Pi_{M0},\Pi_{M1}\}$
($\Pi_{M0}+\Pi_{M1}=1$, $\Pi_{M0}\Pi_{M1}=0$) on the qubit $M$.
After projection $\Pi_{Mi}$, the local state of the qubit $S$
collapses to
\begin{eqnarray}\label{projst}
\rho_{Si}=\frac{1}{\pi_{i}}\left[w|\Psi\rangle_{S}\langle \Psi|p_{i}+
w^{\bot}|\Psi^{\bot}\rangle_{S}\langle\Psi^{\bot}|p_{i}^{\bot}\right.+\nonumber\\
\left.\sqrt{ww^{\bot}}\left(|\Psi\rangle_{S}\langle \Psi^{\bot}|c^{i}+\mbox{h.c.}\right)\right],
\end{eqnarray}
where $\pi_{i}=\mbox{Tr}\rho_{Si}=wp_{i}+w^{\bot}p_{i}^{\bot}$ is the probability of the projection.
Thus after the meter measurement we obtain two sub-ensembles of the states
$\rho_{S0}$ and $\rho_{S1}$ weighted with probabilities $\pi_{0}$ and $\pi_{1}$.
We denote $w_{i}=wp_{i}/\pi_{i}$ and $w^{\bot}_{i}=w^{\bot}p^{\bot}_{i}/\pi_{i}$.

Adopting maximum likelihood estimation strategy, we guess for each
event that the measurement $\Pi_{S}$ gave the most likely results either $0$
or $1$. Our strategy is maximize the likelihood function
$L=\mbox{Max}\{w_0,w_1\}$. The knowledge in a binary decision
problem is the fractional excess of right guesses over wrong
guesses in many experiments repeated under identical conditions
\cite{Englert}. If we have $70\%$ of right guesses and $30\%$ of
wrong guesses than our knowledge is $0.4$. In our task a priori
knowledge without the measurement $\Pi_{M}$ is ${\bf
P}(\Pi_{S})=|w-w^{\bot}|$. A sub-ensemble knowledge after the
particular projection $\Pi_{Mi}$ is ${\bf P}(\Pi_{Mi}\rightarrow
\Pi_{S})=|w_{i}-w_{i}^{\bot}|$. Then after the meter measurement,
an amount of the knowledge ${\bf K}(\Pi_{M}\rightarrow \Pi_{S})$
is the $\pi_{i}$-weighted sum of sub-ensembles knowledges
\cite{Englert}
\begin{eqnarray}\label{know1}
{\bf K}(\Pi_{M}\rightarrow \Pi_{S})&=&
\sum_{i}\pi_{i}{\bf P}(\Pi_{Mi}\rightarrow \Pi_{S})=\nonumber\\
& &\sum_{i}\pi_{i}|w_{i}-w_{i}^{\bot}|,
\end{eqnarray}
where ${\bf P}(\Pi_{S}) \leq {\bf  K}(\Pi_{M}\rightarrow \Pi_{S})\leq 1$. The knowledge
excess ${\bf \Delta K}(\Pi_{M}\rightarrow \Pi_{S})$ is that amount of knowledge which
exceeds the apriori knowledge ${\bf P}(\Pi_{S})=|w-w^{\bot}|$, explicitly
\begin{equation}
{\bf \Delta K}(\Pi_{M}\rightarrow \Pi_{S})={\bf K}(\Pi_{M}\rightarrow \Pi_{S})-{\bf P}(\Pi_{S}),
\end{equation}
where $0\leq {\bf \Delta K}(\Pi_{M}\rightarrow \Pi_{S})\leq 1$.
The knowledge excess quantifies only a part of knowledge which is
gained from the measurement on $M$. If we are not able to extract
any extra knowledge from the measurement on $M$ then the knowledge
excess is vanishing. Using the expansion (\ref{state}) we have
\begin{equation}
{\bf \Delta K}(\Pi_{M}\rightarrow \Pi_{S})=\sum_{i}|\mbox{Tr}_{M}\Pi_{i}(w\rho_{M}-w^{\bot}\rho_{M}^{\bot})|-|w-w^{\bot}|.
\end{equation}
The largest ${\bf \Delta K}(\Pi_{M}\rightarrow \Pi_{S})$ over all $\Pi_{M}$ is the distinguishability excess
${\bf \Delta D}(\Pi_{S})$
\begin{eqnarray}\label{dist}
{\bf \Delta D}(\Pi_{S})=\mbox{Max}_{\Pi_{M}}{\bf \Delta K}(\Pi_{M}\rightarrow \Pi_{S})=\nonumber\\
\mbox{Tr}_{M}|(w\rho_{M}-w^{\bot}\rho_{M}^{\bot})|-|w-w^{\bot}|,
\end{eqnarray}
where $|A|=\sqrt{A^{\dag}A}$,
and thus $0\leq {\bf \Delta K}(\Pi_{M}\rightarrow \Pi_{S})\leq {\bf \Delta D}(\Pi_{S})$.
The analogical quantities ${\bf \Delta K}(\Pi'_{M}\rightarrow \Pi'_{S})$ and
${\bf \Delta D}(\Pi'_{S})$ can be defined for the complementary measurement $\Pi'_{S}$.

Now we shortly prove a relation between the knowledge excesses for $\Pi_{M}=\Pi'_{M}$.
The derivation is inspired by a similar one in Ref.~\cite{Englert}. The sub-ensemble
knowledge about prediction of the complementary projection along the
state $|\Psi'\rangle_{S}=(|\Psi\rangle_{S}+e^{i\theta}|\Psi_{\bot}\rangle_{S})/\sqrt{2}$,
after a particular projection $\Pi_{Mi}$, is ${\bf P}(\Pi_{Mi}\rightarrow \Pi'_{S})=2\sqrt{w_{i}w_{i}^{\bot}}
\mbox{Re}[c_{i}\exp(i\theta)]\leq 2\sqrt{w_{i}w_{i}^{\bot}}|c_{i}|$.
Since $|c_{i}|^{2}\leq 1$ we have
\begin{equation}
{\bf P}^{2}(\Pi_{Mi}\rightarrow \Pi_{S})+{\bf P}^{2}(\Pi_{Mi}\rightarrow \Pi'_{S})\leq 1
\end{equation}
and using Schwarz inequality $(\sum_{k}a_{k}b_{k})^2\leq \sum_{k}a_{k}^{2}
\sum_{l}b_{l}^{2}$, we obtain
\begin{eqnarray}\label{help}
{\bf P}(\Pi_{Mi}\rightarrow \Pi_{S}){\bf P}(\Pi_{Mj}\rightarrow \Pi_{S})+\nonumber\\
{\bf P}(\Pi_{Mi}\rightarrow \Pi'_{S}){\bf P}(\Pi_{Mj}\rightarrow \Pi'_{S})\leq 1.
\end{eqnarray}
Then using (\ref{know1}) and (\ref{help}) we can straightforwardly derive that for $\Pi'_{M}=\Pi_{M}$ the knowledge excesses satisfy
\begin{equation}\label{ineq2}
{\bf \Delta K}^{2}(\Pi_{M}\rightarrow \Pi_{S})+{\bf \Delta K}^{2}(\Pi_{M}\rightarrow \Pi'_{S})\leq 1.
\end{equation}
Using the same
measurement on $M$, the sum of squares of the knowledge excesses
can never overcome unity. It is a duality between the knowledge excesses from a single meter measurement.

\section{Knowledge excesses and Bell-inequality violation}

Accomplishing generally different $\Pi'_{M}$, $\Pi_{M}$, the sum
of squares of the knowledge excesses can be larger than unity if
the state violates the Bell inequalities. Let us discuss in this
case a limitation of the knowledge excesses for any mixed state of the
two-qubit system
\begin{equation}\label{expansion2}
\rho_{SM}=\frac{1}{4}(1\otimes 1+1\otimes\sum_{l=1}^{3} m_{l}\sigma_{l}+
\sum_{l=1}^{3} n_{l}\sigma_{l}\otimes 1+\sum_{k,l=1}^{3}t_{kl}\sigma_{k}\otimes\sigma_{l}),
\end{equation}
where $1$ stands for the identity operator, $m_{l},n_{l}$ are
vectors in $R^{3}$, $\sigma_{l}$ are the standard Pauli operators
and $|V\rangle_{S}$, $|H\rangle_{S}$ are the eigenstates of
$\sigma_{3}$. The coefficients $t_{kl}$ form a real correlation
matrix $T$ and vectors $m_{l}$ and $n_{l}$ determine the local
states $\rho_{S}=\frac{1}{2}(1+n_{l}\sigma_{l})$,
$\rho_{M}=\frac{1}{2}(1+m_{l}\sigma_{l})$. We assume a subset of
states $\bar{\rho}_{SM}$ with the diagonal correlation tensor
$\bar{T}=\mbox{diag}(\bar{t}_{33},\bar{t}_{11},\bar{t}_{22})$,
$\bar{t}_{33}^{2}\geq\bar{t}_{11}^{2},\bar{t}_{22}^{2}$ and with
vectors $\bar{m}_{l}$ and $\bar{n}_{l}$. Any mixed state
$\rho_{SM}$ can be uniquely converted to some $\bar{\rho}_{SM}$
using appropriate local unitary operations. Further we have the
following ordering of the diagonal elements
$\bar{t}_{11}^{2}\geq\bar{t}_{22}^{2}$ or
$\bar{t}_{11}^{2}\leq\bar{t}_{22}^{2}$. Since the prove for
$\bar{t}_{11}^{2}\leq\bar{t}_{22}^{2}$ is only analogical we will
shortly discuss afterward.

According to the strongest correlation $|\bar{t}_{33}|$, one
measurement on $S$ is naturally chosen as
$\bar{\Pi}_{S}=\{|V\rangle_{S}\langle V|,|H\rangle_{S}\langle
H|\}$ and from the expansion (\ref{state}) we obtain using
(\ref{dist}) we have ${\bf \bar{P}}=|n_{3}|$ and either ${\bf
\Delta \bar{D}}=0$ or ${\bf \Delta
\bar{D}}=|\bar{t}_{33}|-|\bar{n}_{3}|>0$. And according to the
second strongest correlation $|\bar{t}_{11}|$, the complementary
measurement is chosen as $\bar{\Pi'}_{S}=\{|+\rangle_{S}\langle
+|,|-\rangle_{S}\langle -|\}$,
$|\pm\rangle_{S}=\frac{1}{\sqrt{2}}(|V\rangle_{S}\pm|H\rangle_{S})$
and then we analogically have ${\bf \bar{P}'}=|n_{1}|$ and either
${\bf \Delta \bar{D}'}=0$ or ${\bf \Delta
\bar{D}'}=|\bar{t}_{11}|-|\bar{n}_{1}|>0$. We can also express a
violation of Bell inequalities in a simple way for such states
with a diagonal $\bar{T}$. Adopting criterion in
Ref.~(\cite{Horodecki95,Zukowski01}), a state $\rho_{SM}$ violates
a Bell inequality if its maximal Bell factor is
\begin{equation}\label{bell1}
2<B_{\mbox{max}}=2\sqrt{\bar{t}_{11}^{2}+\bar{t}_{33}^{2}}\leq 2\sqrt{2}
\end{equation}
since the factor $B_{\mbox{max}}\in \langle 0,2\sqrt{2}\rangle$ is
invariant under local unitary transformations $U_{S}\otimes
U_{M}$ of the state. Similarly, we can derive analogical result
for the case when $\bar{t}_{11}^{2}\leq\bar{t}_{22}^{2}$, either
we have ${\bf \Delta \bar{D}}=0$ or ${\bf \Delta
\bar{D}}=|\bar{t}_{33}|-|\bar{n}_{3}|>0$ and for the complementary
measurement, either ${\bf \Delta D'}=0$ or ${\bf \Delta
D'}=|\bar{t}_{22}|-|\bar{n}_{2}|>0$ and maximum of the Bell factor
is $B_{\mbox{max}}=2\sqrt{\bar{t}_{22}^{2}+\bar{t}_{33}^{2}}$.
Then we can generally obtain that
\begin{equation}
{\bf \Delta \bar{D}}^{2}+{\bf \Delta
\bar{D}}^{'2}\leq\left(\frac{B_{\mbox{max}}}{2}\right)^{2}
\end{equation}
for arbitrary state $\rho_{SM}$ with diagonal $T$, where ${\bf
\Delta \bar{D}},{\bf \Delta \bar{D}}^{'}$ corresponds to the two
largest correlations. The equality is obtained for the states
having vanishing priori knowledge in any basis.

Now we generalize the discussion for any state $\rho_{SM}$
assuming arbitrary measurements
$\Pi_{S},\Pi'_{S},\Pi_{M},\Pi'_{M}$. Any mixed two-qubit state can
be uniquely prepared from the set of states $\bar{\rho}_{SM}$ by
appropriate local unitary transformations $U_{S},U_{M}$ on the
qubits $S$ and $M$. Further, the transformation of the
measurements $\bar{\Pi}_{S}$ and $\bar{\Pi'}_{S}$ to the arbitrary
but still complementary ones effectively corresponds the extra
local unitary transformation $U(\Pi_{S})$ of the state qubit $S$.
Since the distinguishabilities ${\bf \Delta D}(\Pi_{S})$, ${\bf
\Delta D}(\Pi'_{S})$ are generally invariant under any local
unitary transformation on the meter $M$, it is sufficient to
implement a joint unitary transformation
$\tilde{U}_{S}=U(\Pi_{S})U_{S}$ only on the qubit $S$. Further we
can preserve the measurements $\bar{\Pi}_{S}$ and $\bar{\Pi'}_{S}$
in a general prove.

For any unitary transformations $U$ there are unique rotations $O$ such that
$U(\vec{n}.\vec{\sigma})U^{\dag}=(O\vec{n}).\vec{\sigma}$.
If a state $\rho_{SM}$ with diagonal $\bar{T}$ is subjected to the $U_{S}\otimes U_{M}$
transformation its correlation matrix transforms as follows \cite{Horodecki96b}
\begin{equation}
T=O_{S}\bar{T}O_{M}^{\dag}.
\end{equation}
Thus a joint unitary transformation
$\tilde{U}_{S}(\alpha,\beta,\gamma)$ can be represented as a
transformation of the correlation tensor $T=O(\alpha,\beta,\gamma)\bar{T}$,
where $O(\alpha,\beta,\gamma)$ is the matrix of rotation in $R^{3}$ space
\begin{eqnarray}
\left(\begin{array}{ccc}
o_{11} & o_{12} & o_{13}\\
o_{21} & o_{22} & o_{23}\\
o_{31} & o_{32} & o_{33}\\
\end{array}\right)
\end{eqnarray}
with the elements
\begin{eqnarray}\label{elem}
o_{11}&=&\cos\alpha\cos\beta\cos\gamma-\sin\alpha\sin\gamma,\nonumber\\
o_{12}&=& -\cos\alpha\cos\beta\sin\gamma-\sin\alpha\cos\gamma,\nonumber\\
o_{13}&=&\cos\alpha\sin\beta,\nonumber\\
o_{21}&=&\sin\alpha\cos\beta\cos\gamma+\cos\alpha\sin\gamma,\nonumber\\
o_{22}&=&-\sin\alpha\cos\beta\sin\gamma+\cos\alpha
\cos\gamma,\nonumber\\
o_{23}&=&\sin\alpha\sin\beta,\nonumber\\
o_{31}&=&-\sin\beta\cos\gamma,\nonumber\\
o_{32}&=&\sin\beta\sin\gamma,\nonumber\\
o_{33}&=&\cos\beta,\nonumber
\end{eqnarray}
where $\alpha\in \langle 0,2\pi)$, $\beta\in \langle 0,\pi)$, $\gamma\in \langle 0,2\pi)$.

We explicitly calculate ${\bf \Delta D}(\Pi_{S})$ and ${\bf \Delta
D}(\Pi'_{S})$ for the state after the previous transformation.
First we assume that $\bar{t}_{11}^{2}\geq\bar{t}_{22}^{2}$. For
the first measurement from the complementary measurements we
obtain either ${\bf \Delta D}(\Pi_{S})=0$ or
\begin{equation}\label{disc1}
{\bf \Delta D}(\Pi_{S})=\sqrt{t^{2}_{33}+t_{32}^{2}+t_{31}^{2}}-|n_{3}|>0.
\end{equation}
For the second complementary measurement is either ${\bf \Delta D}(\Pi'_{S})=0$ or
\begin{equation}\label{disc2}
{\bf \Delta D}(\Pi'_{S})=\sqrt{t^{2}_{11}+t_{12}^{2}+t_{13}^{2}}-|n_{1}|>0.
\end{equation}
and using the transformation $T=O(\alpha,\beta,\gamma)\bar{T}$ we can derive that
\begin{eqnarray}\label{newD}
{\bf \Delta D}^{2}(\Pi_{S})&\leq&\bar{t}_{33}^{\,2}\cos^{2}\beta+\nonumber\\
& &\sin^{2}\beta(\bar{t}_{11}^{\,2}
\cos^{2}\alpha+\bar{t}_{22}^{\,2}\sin^{2}\alpha),\nonumber\\
{\bf \Delta D}^{2}(\Pi'_{S})&\leq&\bar{t}_{11}^{\,2}(\cos\alpha\cos\beta\cos\gamma-\sin\alpha\sin\gamma)^2+
\nonumber\\
& &\bar{t}_{22}^{\,2}(\cos\alpha\cos\beta\sin\gamma+\sin\alpha\cos\gamma)^2+\nonumber\\
& &\bar{t}_{33}^{\,2}\cos^{2}\alpha\sin^{2}\beta.
\end{eqnarray}
Consequently, assuming
$\bar{t}_{33}^{\,2}>\bar{t}_{11}^{\,2}>\bar{t}_{22}^{\,2}$ we can
prove the following inequality
\begin{eqnarray}
{\bf \Delta D}^{2}(\Pi_{S})+{\bf \Delta D}^{2}(\Pi'_{S})\leq
\bar{t}_{11}^{\,2}+\bar{t}_{33}^{\,2}=\left(\frac{B_{\mbox{max}}}{2}\right)^{2}.
\end{eqnarray}

For $\bar{t}_{11}^{2}\leq\bar{t}_{22}^{2}$, by repeating our
calculation we have for the first measurement the result
(\ref{disc1}) the second complementary measurement is either ${\bf
\Delta D}(\Pi'_{S})=0$ or
\begin{equation}\label{disc3}
{\bf \Delta D}(\Pi'_{S})=\sqrt{t^{2}_{21}+t_{22}^{2}+t_{23}^{2}}-|n_{2}|>0.
\end{equation}
Using (\ref{elem}) we derive
\begin{eqnarray}\label{newD}
{\bf \Delta D}^{2}(\Pi'_{S})&\leq&\bar{t}_{11}^{\,2}(\sin\alpha\cos\beta\cos\gamma+\cos\alpha\sin\gamma)^2+
\nonumber\\
& &\bar{t}_{22}^{\,2}(-\sin\alpha\cos\beta\sin\gamma+\cos\alpha\cos\gamma)^2+\nonumber\\
& &\bar{t}_{33}^{\,2}\sin^{2}\alpha\sin^{2}\beta
\end{eqnarray}
and subsequently,
\begin{eqnarray}
{\bf \Delta D}^{2}(\Pi_{S})+{\bf \Delta D}^{2}(\Pi'_{S})\leq
\bar{t}_{22}^{\,2}+\bar{t}_{33}^{\,2}=\left(\frac{B_{\mbox{max}}}{2}\right)^{2}.
\end{eqnarray}

Finally, since ${\bf \Delta K}(\Pi_{M}\rightarrow\Pi_{S})\leq {\bf
\Delta D}(\Pi_{S})$ and ${\bf \Delta
K}(\Pi'_{M}\rightarrow\Pi'_{S})\leq {\bf \Delta D}(\Pi'_{S})$ we
prove that
\begin{equation}\label{ineq1}
{\bf \Delta K}^{2}(\Pi_{M}\rightarrow\Pi_{S})+{\bf \Delta K}^{2}(\Pi'_{M}\rightarrow\Pi'_{S})\leq
\left(\frac{B_{\mbox{max}}}{2}\right)^{2}
\end{equation}
is generally satisfied. Thus the maximum of Bell factor represents
an important bound on the squares of the excess of knowledge which
can be extracted from the meter measurements. For class of the
states with vanishing priori knowledge for any measurements
$\Pi_{S},\Pi'_{S}$, i.e. $n_{1}=n_{2}=n_{3}=0$, this inequality
can be saturated only by an appropriate choice of the measurements
$\Pi_{S},\Pi'_{S},\Pi_{M},\Pi'_{M}$. For a mixture of Bell states
(\ref{mixbell}) we can find that ${\bf \Delta
D}=|p_{1}-p_{2}+p_{3}-p_{4}|$, ${\bf \Delta
D'}=|p_{1}+p_{2}-p_{3}-p_{4}|$ and
$B_{\mbox{max}}=2\sqrt{2}\sqrt{(p_{1}-p_{4})^{2}+(p_{2}-p_{3})^{2}}$
which saturates the relation (\ref{ineq1}).

It was shown that local filtering operations on single copy of the
state can increase the degree of violation of Bell inequalities
\cite{Gisin96}. There is a unique local (stochastically
reversible) filtering operation $F_{S}$ and $F_{M}$
($F_{S}^{\dag}F_{S}\leq 1_{S}$ and $F_{M}^{\dag}F_{M}\leq 1_{M}$)
on single copy of the state
\begin{equation}
\rho'_{SM}=\frac{F_{S}\otimes F_{M}\rho_{SM}F^{\dag}_{S}\otimes F^{\dag}_{M}}{
\mbox{Tr}(F_{S}\otimes F_{M}\rho_{SM}F^{\dag}_{S}\otimes F^{\dag}_{M})}
\end{equation}
which transforms with some non-zero probability any two-qubit mixed state into a state which is diagonal in Bell basis
\begin{eqnarray}\label{mixbell}
\rho'_{SM}&=&p_{1}|\Psi_{-}\rangle\langle\Psi_{-}|+p_{2}|\Phi_{-}\rangle\langle\Phi_{-}|+\nonumber\\
& &p_{3}|\Psi_{+}\rangle\langle\Psi_{+}|+p_{4}|\Phi_{+}\rangle\langle\Phi_{+}|
\end{eqnarray}
having the largest $B'_{\mbox{max}}\geq B_{\mbox{max}}$
\cite{Verstraete02}. A two-qubit mixed state can be
uniquely bring to this Bell diagonal form with the maximal
violation either with finite probability or asymptotically. For the Bell diagonal states, the Bell
violation cannot be increased by any local filtering on a single
copy. Naturally, after the local filtering the inequality
(\ref{ineq1}) is still satisfied also for the remaining state
$\rho'_{SM}$. The Bell diagonal states have both the local states
maximally disordered, both the apriori knowledges vanish and we can
always saturate the inequality (\ref{ineq1}) with the upper bound
given by $B'_{\mbox{max}}$ only by an appropriate choice of the
measurements $\Pi_{S},\Pi'_{S},\Pi_{M},\Pi'_{M}$. Thus assuming
that $B_{\mbox{max}}$ in the inequality (\ref{ineq1}) is then maximum
under local filtering and unitary operations on a single copy of
the state, the inequality is satisfied for any mixed state and can
be for any mixed state saturated if we use an appropriate local
filtering.

We analyze two interesting and experimentally feasible examples of
the Bell-diagonal states. In both cases, we can use as a source
state the maximally entangled state
$|\Psi_{-}\rangle_{SM}=\frac{1}{\sqrt{2}}(|VH\rangle_{SM}-|HV\rangle_{SM})$
produced by the SPDC process. Evidently, the source state
maximally violates Bell inequalities and has ${\bf \Delta D}={\bf
\Delta D'}=1$. In the first example, we simultaneously perform a
depolarization by (i) random flip of linear polarizations
$|V\rangle_{M}\leftrightarrow |H\rangle_{M}$ with the probability
$(1-R_{1})/2$ and simultaneously, by (ii) phase-shift $\pi$
between linear polarizations $|V\rangle_{M}\rightarrow
|V\rangle_{M}$, $|H\rangle_{M}\rightarrow -|H\rangle_{M}$ with the
probability $(1-R_{2})/2$. As a result of the depolarizing
procedure we prepare a mixture of Bell states
\begin{eqnarray}\label{state1}
\rho_{1}&=&\frac{1+R_{2}}{2}(\frac{1+R_{1}}{2}|\Psi_{-}\rangle\langle\Psi_{-}
|+\frac{1-R_{1}}{2}|\Phi_{-}\rangle\langle\Phi_{-}|)+\nonumber\\
& &\frac{1-R_{2}}{2}(\frac{1+R_{1}}{2}|\Psi_{+}\rangle\langle\Psi_{+}|+\frac{1-R_{1}}{2}|\Phi_{+}\rangle\langle\Phi_{+}|).\nonumber\\
\end{eqnarray}
For this case, the distinguishability excess is ${\bf \Delta
D}=R_{1}$ and vanishes only if the random polarization flip has
larger probability, whereas the complementary distinguishability
excess is ${\bf \Delta D'}=R_{2}$ and decreases only with an
increasing probability of the random polarization phase-shift. The
maximal value of Bell factor is
$B_{\mbox{max}}=2\sqrt{R_{1}^{2}+R_{2}^{2}}$ and the inequality
(\ref{ineq1}) is saturated. Thus we can independently control both
the complementary distinguishability excesses. For $R_{1}=1$ ($R_{2}=1$)
and changing $R_{2}$ ($R_{1}$) we are able to extract the maximal
unit distinguishability excess ${\bf \Delta D}$ (${\bf \Delta
D'}$) irrespective to the complementary one ${\bf \Delta D'}$
(${\bf \Delta D}$). For $R_{1}=0$ ($R_{2}=0$) and controlling
$R_{2}$ ($R_{1}$), the depolarization prevents us to extract any
knowledge excess since ${\bf \Delta D}=0$ (${\bf \Delta D'}=0$)
however we are still able to obtain the complementary knowledge
excess ${\bf \Delta D'}>0$ (${\bf \Delta D'}>0$).

In the second example, we show as both ${\bf \Delta D}$ and ${\bf
\Delta D'}$ can be gradually enhanced sharing separable state,
entangled state which satisfies the Bell inequalities and
consequently, sharing entangled state violating Bell inequalities.
We assume a loss of both the knowledge excesses from the state
$|\Psi_{-}\rangle_{SM}$ by an extraction of the photon in the
meter beam with probability $1-R$ and its substitution by another
photon with a completely random polarization. Thus we detect the
results produced by a mixture of the Bell states which is known as
Werner state \cite{Werner}
\begin{equation}\label{state2}
\rho_2=R|\Psi_{-}\rangle\langle\Psi_{-}|+\frac{1-R}{4}1\otimes 1.
\end{equation}
In this case, a loss of the distinguishability excesses with
decreasing $R$ are identical ${\bf \Delta D}={\bf \Delta D'}=R$.
We know that the Werner's state is entangled only for $R>1/3$ and
violates Bell inequalities, having maximum of Bell factor
$B_{\mbox{max}}=2\sqrt{2}R$, only if $R>1/\sqrt{2}$. Also in this
case, the inequality (\ref{ineq1}) can be saturated. Since we have
an entangled state non-violating Bell inequalities for $1/3 \leq
R\leq 1/\sqrt{2}$ and non-entangled state for $R\leq 1/3$ it means
that we can observe both non-vanishing distinguishability excesses ${\bf \Delta D}
< 1$ and ${\bf \Delta D'} < 1$ even in a case of classical
correlated states.

\medskip
\noindent {\bf Acknowledgments} We would like to thank J. Fiur\'
a\v sek, L. Mi\v sta Jr. for the fruitful discussions. The work
was supported by the projects 202/03/D239 of GACR and LN00A015 and
CEZ: J14/98 of the Ministry of Education of Czech Republic.



\begin{thebibliography}{99}
\bibitem{Bell}
Theoretically, for a given state $\rho$
maximal value of the Bell factor $B$ can be theoretically calculated using the
following formula $B_{\mbox{max}}=2\sqrt{M(\rho)}$, where
$M(\rho)$ is real-valued function of the density matrix
$\rho$ \cite{Horodecki95}.
To define $M(\rho)$, one needs a $3\times3$ matrix $T$ with
elements $T_{ij}=\mbox{Tr}\{\rho (\sigma_{i}\otimes\sigma_{j})\}$. Then,
$M(\rho)$ is the sum of the two largest eigenvalues of the Hermitian
matrix $T^{\dag}T$. If $B_{\mbox{max}}>2$ then Bell inequalitie are violated and
all experimental outcomes which state $\rho$ generate cannot be explained only by local realism and
quantum theory must be used \cite{Zukowski01}.
\end{thebibliography}
\end{document}